\documentclass[a4paper,12pt]{article}

\usepackage{psfrag}
\usepackage{mathrsfs} %\mathscr

\usepackage{graphicx}
\usepackage{subfigure}
\usepackage[margin=10pt,font=small,labelfont=bf,
labelsep=period]{caption}
\usepackage{amsmath}
\usepackage{amssymb}
\usepackage{dsfont} %\mathds
\usepackage[bottom]{footmisc} %footnotes always at bottom of page
\usepackage{indentfirst} %indent the first line of sections
\usepackage{upref} % upright reference numbers in italics environment
\usepackage{cite} % multiple citations e.g. [16-22]
\numberwithin{equation}{section}

\renewcommand{\baselinestretch}{1.5}
\begin{document}
\begin{titlepage}
\renewcommand{\baselinestretch}{1.1}
\title{\begin{flushright}
\normalsize{MZ-TH/11-01}
\bigskip
\vspace{1cm}
\end{flushright}
QED coupled to QEG}
\date{}
\author{U. Harst and M. Reuter\\
{\small Institute of Physics, University of Mainz}\\[-0.2cm]
{\small Staudingerweg 7, D-55099 Mainz, Germany}}
\maketitle\thispagestyle{empty}
\begin{abstract} 
We discuss the non-perturbative renormalization group flow of Quantum 
Electrodynamics (QED) coupled to Quantum Einstein Gravity (QEG) and explore the 
possibilities for defining its continuum limit at a fixed point that would lead 
to a non-trivial, i.\,e. interacting field theory. We find two fixed points 
suitable for the Asymptotic Safety construction. In the first case, the 
fine-structure constant $\alpha$ vanishes at the fixed point and its infrared 
(``renormalized'') value is a free parameter not determined by the theory 
itself. In the second case, the fixed point value of $\alpha$ is non-zero, and 
its infrared value is a computable prediction of the theory.
\end{abstract}
\end{titlepage}
\newpage 
\section{Introduction}
With the advent of perturbative renormalization theory in the late fourties of 
the last century Quantum Electrodynamics (QED) matured to a physical theory of 
unprecedented predictive power. With only two input parameters, the electron's 
mass and charge, it is able to describe, or ``explain'' a wealth of 
experimental data, often with spectacular precision. Later on, in the 
seventies, the electromagnetic, weak and strong interactions were united in the 
broader but conceptually similar framework of what is now known as the standard 
model of elementary particle physics. Undoubtedly the standard model provides a 
very impressive description of all three interactions but it also highlights 
the very limitations of quantum field theory in its familiar form, namely the 
fact that there is always a set of parameters (masses, couplings, mixing 
angles, etc.) which, as a matter of principle, cannot be computed within the 
theory and must be extracted from the experiment. In QED this concerns only the 
electron's mass and charge, but in the standard model there are already more 
than two dozens of similar input parameters. In fact, while in the construction 
of the standard model non-abelian gauge fields and spontaneous symmetry 
breaking made their appearance as new ingredients, the way the pertinent 
quantum field theory is ``defined'', i.\,e. how the infinite cutoff limit is 
taken, remained essentially the same as in QED. Using K. Wilson's modern 
picture of renormalization \cite{wilson} the procedure of perturbative 
renormalization can be viewed as taking the infinite cutoff limit at a trivial, 
or Gaussian fixed point of the renormalization group flow. The dimensionality 
of its critical manifold decides about the number of undetermined input 
parameters. Since the fixed point is Gaussian, the corresponding scaling 
dimensions are essentially the canonical ones, and as a result the generalized 
couplings that appear as coefficients of field monomials with a mass dimension 
not larger than 4 are essentially the ones which cannot be computed. In QED 
these are the monomials $\bar{\psi} \psi$ and $\bar{\psi} \gamma^\mu \psi 
A_\mu$ with canonical dimensions 2 and 4, respectively.

If one takes the infinite cutoff limit of some quantum field theory at a 
non-trivial, or non-Gaussian fixed point \cite{parisi} the corresponding count 
relies on the actual scaling dimensions given by the renormalization group (RG) 
flow linearized about this particular non-Gaussian fixed point (NGFP). Those 
scaling dimensions will in general not coincide with the canonical mass 
dimension of some field monomial. Nevertheless, as we shall see in a moment, 
the number $s$ of parameters which are undetermined in principle is 
again given by the stability (attractivity) properties of the fixed point. It 
equals the dimensionality of its ultraviolet critical hypersurface 
$\mathscr{S}_{\text{UV}}$. By definition, $\mathscr{S}_{\text{UV}}$ is the 
set of all points in coupling constant space which is mapped onto the fixed 
point by the inverse\footnote{In this paper the orientation of the RG flow is 
always from the ultraviolet (UV) towards the infrared (IR), i.\,e. in the 
direction of the natural coarse graining flow.} RG flow.

Computing the set of all scaling dimensions at a NGFP, provided it exists, is a 
difficult task which requires hard (non-perturbative) computations, in contrast 
to the Gaussian case where (leaving marginal cases aside) pure power counting 
gives the essential picture.

Therefore, in principle it is conceivable that if QED or the entire standard 
model, say, should possess a NGFP appropriate for taking the infinite cutoff 
limit the resulting number of undetermined parameters, $s= 
\text{dim}\, \mathscr{S}_{\text{UV}}$, turns out smaller than at the 
trivial fixed point. If so, the quantum theory based upon the NGFP would have a 
higher degree of predictivity than its ``Gaussian'' counterpart which is 
provided by the familiar perturbative renormalization procedure. A theory 
defined by means of a NGFP is sometimes referred to as ``non-perturbatively 
renormalized''. In this paper we shall encounter a version of QED which could 
possess such a higher degree of predictivity.

Already in the early years of QED a non-trivial UV fixed point was speculated 
about, for a somewhat different reason though \cite{gell}. The well-known one 
loop formula
\begin{equation}\label{eRenormalized}
 \frac{1}{e_{\text{ren}}^2}-\frac{1}{e_\Lambda^2}
 =
 \frac{1}{6 \pi^2}\ln\left(\frac{\Lambda}{m_{\text{ren}}}\right)
\end{equation}
suggests that it might be difficult to obtain an {\it interacting} theory in 
the limit when the UV-cutoff $\Lambda$ is removed: Keeping in 
\eqref{eRenormalized} the value $e_\Lambda$ of the bare charge fixed and 
sending $\Lambda$ to infinity one finds that the renormalized charge 
$e_\text{ren}$ vanishes, so one is left with a ``trivial'' theory. 
(Conversely, keeping $e_\text{ren}$ fixed, it is impossible to let $\Lambda 
\rightarrow \infty$ since $e_\Lambda$ diverges at a finite value of $\Lambda$, 
the Landau pole.) Clearly, if the exact version of eq.\,\eqref{eRenormalized} 
displays a UV fixed point such that $e_\Lambda \rightarrow e^*$ for $\Lambda 
\rightarrow \infty$ the prospects for an interacting, cutoff-free theory were 
much better.

However, to the best of our knowledge no such fixed point exists. On the 
contrary, comprehensive lattice simulations \cite{lattice} and studies using 
non-perturbative functional RG methods \cite{gies-QED} lead to the conclusion 
that QED is very likely to be a trivial theory in 4 dimensions.

As we shall argue later on, the situation might be different when QED is 
coupled to quantized gravity.

Trying to take the infinite cutoff limit at a NGFP, provided there exists any, 
is an option even when the theory under consideration is non-renormalizable in 
perturbation theory. A well known example of a perturbatively 
non-renormalizable theory which can be defined in this way (which is 
``non-perturbatively renormalizable'') is the Gross-Neveu model in 3 dimensions 
\cite{parisi}.

The situation seems to be similar in Einstein gravity in 4 dimensions. While 
the perturbative quantization of general relativity leads to a 
non-renormalizable theory \cite{tHooft-Velt, goroff-sag,vandeVen} it now 
appears quite likely that there exists a NGFP suitable for taking the infinite 
cutoff limit there, leading to a predictive theory with only a small number of 
free parameters. The idea of defining Quantum Einstein Gravity (QEG) at a 
non-trivial fixed point is due to S. Weinberg \cite{wein} who coined the term 
Asymptotic Safety for this scenario. This term highlights the analogy with 
asymptotic freedom, the key difference being that now a non-Gaussian rather 
than Gaussian fixed point controls the ultraviolet behavior. While originally 
the viability of the Asymptotic Safety idea could be tested near 2 dimensions 
only \cite{wein}, the gravitational average action introduced later \cite{mr} 
opened the way for detailed non-perturbative studies of QEG in 4 dimensions 
\cite{mr, oliver1, frank1, oliver2,perper1, codelloannphys, franketal, 
reviews}. Besides finding significant evidence for the very existence of a NGFP 
it was also realized that, consistent with general expectations \cite{wein}, 
the dimensionality $s=\text{dim}\, \mathscr{S}_{\text{UV}}$ 
indeed seems to be a small number and probably does not increase beyond a 
certain number when the total dimensionality of the truncated coupling constant 
space is increased.

This phenomenon was first observed in a 3 parameter RG flow in $d$ dimensions 
which included the running Newton constant, cosmological constant and the 
prefactor of a third invariant, $\int d^{\text{d}}x \sqrt{g} R^2$, 
\cite{oliver2}. The flow equations depend on the real (not necessarily integer) 
number $d$ in a continuous way, and it was found that there exists a NGFP for 
every dimensionality in the interval between\footnote{Note, however that no 
$\epsilon$-expansion is performed here \cite{oliver2}.} $d=2+\epsilon$ and 
$d=4$ (at least). Interestingly, its attractivity properties depend on $d$ in 
an essential way: Above a certain critical dimension (near 3), in particular in 
$d=4$, the NGFP is UV attractive in all 3 directions, while below the critical 
dimension one of the three directions is UV repulsive. In particular in 
$d=2+\varepsilon$ it was found that $s=2\,$! Thus the RG 
trajectories of all asymptotically safe theories are confined to a 
2-dimensional surface imbedded in the 3-dimensional parameter space. Hence one 
of the couplings can be predicted in terms of the other two.

The situation is similar in $d=4$ where, however, the ``stabilization'' of 
$s$ at a small value is seen only when a larger parameter (or 
``theory'') space is used. Making an ansatz of the form $\int d^4 x \sqrt{g}\, 
f(R)$ where $f$ is a polynomial in the curvature scalar it was found that 
$\text{dim}\, \mathscr{S}_{\text{UV}}$ stabilized at $s =3$ 
when the degree of the polynomial was increased \cite{codelloannphys}. In a 
calculation with 8 free parameters in $f$ this allowed for 5 predictions in 
terms of 3 input parameters.

In this paper we are going to discuss the issue of non-perturbative 
renormalizability and the possibility of enhanced predictivity for QED coupled 
to quantum gravity. One of the motivations are various recent perturbative 
calculations of quantum gravity corrections to the QED or Yang-Mills beta 
function governing the RG running of the gauge coupling 
\cite{ymg-robwil,ymg-robthesis,ymg-piet,ymg-toms,ymg-ebert,ymg-tangwu,ymg-toms2,
tomsnature,donoghue2010,ellis2010}. The to date most complete perturbative 
analysis uses a gauge fixing independent approach and a regulator which retains 
power-like divergences \cite{tomsnature}. It leads to the following 1-loop RG 
equation for the running of the electric charge with the energy scale $E$:
\begin{equation}\label{1loopToms}
 E \, \frac{\text{d} e (E)}{\text{d} E}
 =
 \frac{e^3}{12 \pi^2} -\frac{e}{\pi} \left(G E^2 +\frac{3}{2} G \Lambda\right)
\end{equation}
The first term on the RHS of \eqref{1loopToms} is the familiar one from the 
fermion loops which tends to increase $e$ at large energies. The second term, 
the gravity correction, involves Newton's constant $G$ and the cosmological 
constant $\Lambda$. It has a negative sign and tries to drive $e$ to smaller 
values as $E$ increases. In fact, it has been claimed \cite{tomsnature} that 
the electric charge vanishes at high energies and may be regarded an {\it 
asymptotically free} coupling therefore.

In the following we shall reconsider this picture in the light of 
asymptotically safe gravity. We shall demonstrate that if QED coupled to QEG is 
asymptotically safe, there exists a second option for the behavior of the 
electric charge at high energies: it may assume a non-zero fixed point value 
$e^*\neq 0$. If this option is realized the asymptotic behavior of QED + QEG is 
governed by a non-Gaussian fixed point whose hypersurface 
$\mathscr{S}_{\text{UV}}$ is likely to have a lower dimension than in the 
corresponding asymptotically {\it free} case $e^*=0$. Within a simple 
truncation of theory space, we find that for the theory with $e^*\neq0$ the 
infrared value of the charge, or the fine-structure constant $\alpha \equiv 
e^2/(4 \pi)$, is a {\it computable number} which is completely fixed by the 
electron mass in Planck units.

The remaining sections of this paper are organized as follows. In Section 2 we 
introduce and motivate the RG equations we are going to study, and in Section 3 
we show that they possess two distinct non-Gaussian fixed points. In Section 4 
we solve a simplified version of the RG equations analytically and in Section 5 
we discuss the Asymptotic Safety scenario related to one of the fixed points 
where the fine-structure constant can be predicted. In Section 6 we supplement 
the investigation by a numerical analysis of the RG flow, and the Conclusions 
will be given in Section 7.

\section{The RG equations \label{RGEq}}
We shall use a projected form of the gravitational average action \cite{mr} to 
describe the non-perturbative RG behavior of QED coupled to QEG in terms of a 
simple 3-dimensional theory space, treating the charge $e(k)$, or equivalently 
the fine-structure constant $\alpha(k)\equiv e(k)^2/(4 \pi)$, along with 
Newton's constant and the cosmological constant as running quantities. 
Combining the results of \cite{mr} for pure gravity in the Einstein-Hilbert 
truncation with the findings of \cite{dhr1,dhr2} for the gravity corrections to 
the running of $\alpha$ we are led to the following ``caricature'' of the flow 
equations:
\begin{subequations}
 \label{FlowEquations}
 \begin{align}
 \partial_t g       &= \beta_g \equiv [2 + \eta_N(g,\lambda)]g    \label{FEg}\\
 \partial_t \lambda &= \beta_\lambda(g,\lambda)                   \label{FEl}\\
 \partial_t \alpha  &= \beta_\alpha \equiv \left(A h_2(\alpha)
                      -\frac{6}{\pi}\Phi^1_1(0) g\right) \alpha   \label{FEa}
 \end{align}
\end{subequations}
with the coefficient
\begin{equation}
 A\equiv\frac{2}{3\pi} n_F.
\end{equation}
Here we consider for illustrative purposes a variant of quantum electrodynamics 
with $n_F$ ``flavors'' of electrons.\\
Several comments are in order now.\\
{\bf (A)} The equations are written in terms of the dimensionless running 
couplings $g(k) \equiv k^2 \, G(k)$, $\lambda(k) \equiv \Lambda(k)/k^2$ and 
$\alpha(k)$ where $k$ is the IR cutoff built into the average action. The 
dimensionless ``RG time'' is denoted $t\equiv \ln(k/k_0).$\\
{\bf (B)} The first two equations, \eqref{FEg} and \eqref{FEl}, are taken to be 
those of pure gravity in the Einstein-Hilbert truncation. The anomalous 
dimension of Newton's constant, $\eta_N(g,\lambda)$, and the beta function for 
the cosmological constant, $\beta_\lambda(g,\lambda)$, were found in ref. 
\cite{mr}.\footnote{For the explicit formulae see eqs.\,(4.41) and (4.43) in 
ref. \cite{mr}.} Neglecting the backreaction of the matter fields on the 
renormalization in the gravity sector is (at least partially) justified by the 
investigations in \cite{perper1} where is was found that a Maxwell field and 
one or a few Dirac fields do not qualitatively alter the RG flow of $g$ and 
$\lambda$; this calculation had assumed free matter fields though.\\
{\bf (C)} For small $g$ the anomalous dimension $\eta_N$ can be expanded in a 
power series in the Newton constant according to
\begin{equation}
 \eta_N
  = B_1(\lambda) h_1(\lambda,g)
  = B_1(\lambda) \left(g + B_2(\lambda)g^2 + \ldots\right)
\end{equation}
with functions $B_1$ and $B_2$ given in \cite{mr}\footnote{See eqs.\,(4.40) in 
ref. \cite{mr}.}. From several non-perturbative calculations \cite{reviews} we 
know the function $h_1(g,\lambda)$ rather precisely and we find that 
$B_1(\lambda)<0$ for all $\lambda$. Those calculations show in particular that 
the running of $g(k)$ does not change very much if one approximates $\lambda(k) 
\approx 0$, $B_1(\lambda)\approx B_1(0), B_2(\lambda)\approx 0$, whence
\begin{equation}
 \eta_N(g,\lambda)\approx B_1(0) g < 0.
\end{equation}
In terms of the standard threshold functions $\Phi^n_p(w)$\footnote{See 
eqs.\,(4.32) in ref. \cite{mr}.} used in \cite{mr} we have explicitly $B_1(0)= 
-1/(3 \pi) \left[ 24 \Phi^2_2(0) - \Phi^1_1(0)\right]$.\\
{\bf (D)} The beta function for $\alpha$ in the third equation, 
eq.\,\eqref{FEa}, involves a pure matter contribution, written as $A 
\,h_2(\alpha)\,\alpha$, and the gravity correction $\propto g$ taken from 
\cite{dhr1}. The former has been computed in perturbation theory, the first two 
terms being (for $n_F=1$)
\begin{equation}
 \left. \beta_\alpha (\alpha)\right|_{g=0} \equiv A \, h_2(\alpha)\, \alpha
 =
 \alpha \left[\frac{2}{3} \left(\frac{\alpha}{\pi}\right)
 +\frac{1}{2}\left(\frac{\alpha}{\pi}\right)^2 + \mathcal{O}(\alpha^3)\right].
\end{equation}
To obtain a qualitative understanding it will be sufficient to employ the 
1-loop approximation 
\begin{equation}
 h_2(\alpha)=\alpha.
\end{equation}
Indeed, the lattice and flow equation studies mentioned above indicate that 
there exists no non-trivial continuum limit for QED (without gravity), and this 
means that $\left. \beta_\alpha(\alpha)\right|_{g=0}$ has no zero at any 
$\alpha>0$. Therefore $h_2(\alpha)= \beta_\alpha/(A \alpha)$ starts out as 
$h_2(\alpha)=\alpha$ in the perturbative regime $\alpha \lesssim 1$, and for 
larger $\alpha$ it is still known to be an increasing function: 
$h_2'(\alpha)>0$. To be able to solve the RG equations analytically we shall 
set $h_2(\alpha)=\alpha$ for all values of $\alpha$. This is a qualitatively 
reliable approximation since, as we shall see, at most a zero of $h_2(\alpha)$ 
could change the general picture.\\
{\bf (E)} As it stands, $\beta_\alpha$ applies only above the threshold due to 
the mass of the electron at $k=m_e$. At $k \lesssim m_e$ the fermion loops no 
longer renormalize $\alpha$. In the full fledged average action formalism this 
decoupling is described by a certain threshold function. Here a simplified 
description will be sufficient where we set $A=0$ if $k<m_e$.\\
{\bf (F)} The gravity contribution on the RHS of \eqref{FEa} was derived in 
ref. \cite{dhr1} within a truncation of theory space which included the gauge 
field action $\frac{1}{4 e^2(k)} \int \text{d}^d x \sqrt{g} F_{\mu \nu} 
F^{\mu \nu}$ besides the Einstein-Hilbert terms. Within the approximation 
considered there the gravity correction to $\partial_t e$ is seen to be 
independent of the interactions within the matter sector, if any. Therefore it 
is the same for QED and the non-abelian Yang-Mills field considered in 
\cite{dhr1} so that we may obtain \eqref{FEa} by simply replacing there the 
non-abelian gauge boson contribution with the corresponding fermion term (of 
the opposite sign!). In \cite{dhr1} also subleading corrections to $\partial_t 
e$ involving the cosmological constant were found. They, too, within their 
domain of reliability do not change the qualitative picture and are omitted 
therefore.\\
{\bf (G)} Identifying the scales $E$ and $k$, the two terms inside the brackets 
on the RHS of the perturbative result \eqref{1loopToms}, in our notation, 
translate to $g+\frac{3}{2} g \lambda$. Thus, for $\lambda=0$, the perturbative 
gravitational correction has the same structure as \eqref{FEa} from the average 
action. Since $\lambda$ is small in the applications below, subleading 
corrections such as the term $\frac{3}{2} g \lambda$ are inessential for the 
qualitative properties of the flow.

We close this section with a word of warning. There is considerable debate in 
the literature about the gravitational corrections to the beta function of 
gauge couplings, on their precise form \cite{ymg-robwil, ymg-robthesis, 
ymg-piet, ymg-toms, ymg-ebert, ymg-tangwu, ymg-toms2, tomsnature}, 
as well as their usefulness \cite{donoghue2010} and observability 
\cite{ellis2010} in scattering experiments. We emphasize that within the 
Asymptotic Safety program these are secondary issues which are not (yet) 
relevant. Our goal is first of all to {\it construct} a quantum field theory by 
devising a way to take the infinite cutoff limit of the corresponding 
functional integral; we do this by replacing the functional integral 
computation by the task of solving an exact RG equation for the effective 
average action $\Gamma_k$ and trying to take the continuum limit at a fixed 
point of its flow. Only once this is achieved one can start to {\it analyze and 
interpret} the resulting theory, and only then questions such as those above on 
scattering experiments can (and should) be asked.

For the time being we are still in the first phase, and so the RG equations 
used in this paper should be seen as a tool towards understanding the flow of 
$\Gamma_k$ and the possible continuum limits it might hint at.

In view of the non-universality of the gravitational corrections \cite{dhr1, 
dhr2} it is also important to stress that a priori all our results hold true 
only for the very definition of $e(k)$ used here, namely via the prefactor of 
the $\int F^2$-term in $\Gamma_k$. Every comparison with different definitions 
in other settings or schemes would require a separate analysis.

\section{The fixed points}
Let us start the analysis of the system \eqref{FlowEquations} by finding its 
fixed points, i.\,e. common zeros $(g^*,\lambda^*,\alpha^*)$ of all three beta 
functions. Obviously there is a trivial or Gaussian fixed point {\bf GFP} at 
$g^*=\lambda^*=\alpha^*=0$.

Furthermore we know that the subsystem of flow equations for pure gravity in 
the Einstein-Hilbert truncation, eqs.\,(\ref{FEg}, \ref{FEl}), admits for a 
NGFP at $(g^*_0,\lambda^*_0)\neq 0$. This fixed point lifts to a NGFP of the 
full system located at $(g^*_0,\lambda^*_0,\alpha^*=0)$. We will denote it by 
{\bf NGFP}${}_1$. This fixed point is trivial from the QED perspective, the 
electromagnetic interaction is ``switched off'' there, while the gravitational 
selfinteraction is the same as in pure gravity.

There exists a second NGFP, non-trivial also in the QED sense, if the equation
\begin{equation}
 h_2(\alpha)=\frac{6}{\pi} \frac{g^*}{A} \Phi^1_1(0)
\end{equation}
has a solution for some $\alpha=\alpha^*\neq 0$. We shall see in a moment that 
this is indeed the case. This fixed point we will call {\bf NGFP}${}_2$.

In the following we will be particularly interested in an Asymptotic Safety 
scenario with respect to this NGFP. Near the fixed point the (linearized) flow 
is governed by the stability matrix $B_{ij}=\partial_{u_j} \beta_{u_i}(u^*)$ 
according to
\begin{equation}
 \partial_t u_i(k) =\sum_{j} B_{ij} \left(u_j(k)-u_j^*\right),
\end{equation}
where $u=(g, \lambda, \alpha)$. For the system under consideration the 
stability matrix is of the form
\begin{equation}
 B=\left. \begin{pmatrix}
    \partial_g \beta_g       & \partial_\lambda \beta_g        & 0 \\
    \partial_g \beta_\lambda & \partial_\lambda \beta_\lambda  & 0 \\
    \partial_g \beta_\alpha  & \partial_\lambda \beta_\alpha   & 
    \partial_\alpha \beta_\alpha
   \end{pmatrix}\right|_{u=u^*}
\end{equation}
Two of its eigenvalues are therefore identical to the case of pure gravity in 
the Einstein-Hilbert truncation, giving rise to the familiar two {\it UV 
attractive} directions \cite{frank1, oliver1}. The third eigenvalue is given by 
\begin{equation}
 \partial_\alpha \beta_\alpha(u^*) = \left( A h_2(\alpha^*)-\frac{6}{\pi} 
 \Phi^1_1(0) g^* \right) + A h_2'(\alpha^*) \alpha^*= A h_2'(\alpha^*) \alpha^*.
\end{equation}
If $\alpha^*$ is positive, which will actually turn out to be the case, the 
sign of $\partial_\alpha \beta_\alpha(u^*)$ agrees with the sign of 
$h_2'(\alpha^*)$.

At this point we take advantage of the information from the lattice and flow 
equations studies trying to find a non-trivial continuum limit of QED without 
gravity. In {\bf(D)} of Section \ref{RGEq} we saw that their negative results 
suggest that $h_2'(\alpha^*)>0$ holds true even beyond perturbation theory. As 
a consequence, the third eigenvalue $\partial_\alpha \beta_\alpha(u^*)$, 
corresponding to the $\alpha$ direction in the 3-dimensional $g$-$\lambda$-$ 
\alpha$--theory space, is {\it UV repulsive}. With two UV attractive and one 
repulsive direction the UV critical hypersurface $\mathscr{S}_{\text{UV}}$ 
pertaining to {\bf NGFP}${}_2$ is a two-dimensional surface in a 3-dimensional 
space, i.\,e. $s_2\,= \text{dim}\, 
\mathscr{S}_{\text{UV}}({\bf NGFP}_2)=2.$

In comparison, let us also analyze the eigenvalues of the stability matrix of 
the other fixed point {\bf NGFP}${}_1$. As $\beta_g$ and $\beta_\lambda$ do not 
depend on $\alpha$ in our approximation the first two eigenvalues remain the 
same as for {\bf NGFP}${}_2$. However, for the third eigenvalue we obtain
\begin{equation}
 \partial_\alpha \beta_\alpha(u^*) =
 \left( A h_2(\alpha^*)-\frac{6}{\pi} \Phi^1_1(0) g^* \right)+ A h_2'(\alpha^*) 
 \alpha^* \stackrel{\alpha^*=0}{=}-\frac{6}{\pi} \Phi^1_1(0) g^*<0,
\end{equation}
such that the third direction turns out to be {\it UV attractive} as well. 
Hence, {\bf NGFP}${}_1$ has a 3-dimensional UV critical hypersurface, i.\,e. 
$s_1\,= \text{dim}\, \mathscr{S}_{\text{UV}}({\bf 
NGFP}_1)=3.$ The fact that $s_2<s_1$ reflects the 
enhanced predictivity of an Asymptotic Safety scenario with respect to {\bf 
NGFP}${}_2$ compared to {\bf NGFP}${}_1$.

\section{Explicit RG trajectories}
Let us now analyze the flow in a simple analytically tractable approximation. 
For that we expand the functions $h_1$ and $h_2$ to first order in $g$ and 
$\alpha$, respectively,
\begin{equation}\label{Approximation}
 h_1(g)= g + \mathcal{O}(g^2) \qquad \text{and} \qquad h_2(\alpha)
       = \alpha + \mathcal{O}(\alpha^2).
\end{equation}
Furthermore, we neglect the running of the cosmological constant and fix 
$\lambda=\lambda_0$ to a constant value. The remaining system of flow equations 
reads
\begin{subequations}\label{FEapprox}
\begin{align}
 \partial_t g &= \left[2+B_1(\lambda_0)\,g\right]g \label{FEgapprox}\\
 \partial_t \alpha &= \left(A \alpha -\frac{6}{\pi}\Phi^1_1(0)\, g 
 \right)\alpha. \label{FEaapprox}
\end{align}
\end{subequations}
In this approximation there clearly exists a ${\bf NGFP}_2$ with fixed point 
values
\begin{equation}\label{FPValues}
 g^*=-\frac{2}{B_1(\lambda_0)} \qquad \text{and} \qquad \alpha^*= 
 \frac{6}{\pi} \frac{g^*}{A} \Phi^1_1(0). 
\end{equation}
In the following we will express the constant $B_1(\lambda_0)$ in terms of the 
fixed point value $g^*$ according to $B_1(\lambda_0)=-2/g^*$.

The approximation allows us to solve \eqref{FEgapprox} in separation; its 
solution is given by
\begin{equation}
 g(k)= \frac{G_0 k^2}{1+\frac{G_0 k^2}{g^*}}\label{gTrajectory}.
\end{equation}
The constant of integration $G_0\equiv \lim_{k \rightarrow 0} g(k)/k^2$ can be 
interpreted as the IR value of the running Newton constant. The simple RG 
trajectory \eqref{gTrajectory} for $g$ shares a crucial feature with any 
asymptotically safe trajectory of the exact system for pure gravity, namely 
that it connects the classical regime $g(k) \approx G_0 k^2$ for $k \ll 
m_\text{Pl} \equiv G_0^{-1/2}$ and the fixed point regime $g(k) \approx g^*$ 
for $k \gg m_\text{Pl}$. Note that the Planck mass is defined in terms of the 
constant $G_0$.

Due to the simplified form of \eqref{FEaapprox}, the RG equation for $\alpha$ 
is now an ordinary differential equation of Riccati type, which can therefore 
be solved in closed form without the need for a specification of the function 
$g(k)$. Its general solution reads, with $\Phi^1_1 \equiv \Phi^1_1(0)$,
\begin{equation}\label{aTrajectory}
 \frac{1}{\alpha(k)}=\frac{1}{\alpha_0} \exp \left(\frac{6}{\pi} \Phi^1_1 
 \int_{k_0}^k \frac{g(k')}{k'}\text{d}k'\right)- A \int_{k_0}^k 
 \exp\left(\frac{6}{\pi} \Phi^1_1 \int_{k'}^{k} 
 \frac{g(k'')}{k''}\text{d}k''\right) \frac{\text{d}k'}{k'},
\end{equation}
where $\alpha_0=\alpha(k_0)$ is the value of the fine-structure constant at a 
fixed reference scale $k_0$. If we now specialize for the function $g(k)$ of 
eq.\,\eqref{gTrajectory} we can perform the integrations in \eqref{aTrajectory} 
and we find
\begin{equation}\label{aTrajectory2}
 \begin{split}
  \frac{1}{\alpha(k)}= \left(\frac{g^*+G_0 k^2}{g^* +G_0 
  k_0^2}\right)^{\frac{3}{\pi} \Phi^1_1 g^*} \left[\frac{1}{\alpha_0} -  
  \frac{1}{\alpha^*} \left( 1\! +\! \frac{g^*}{G_0 k_0^2}\right) {}_2 
  F_1\left(\!1,1,1\!+\!\frac{3}{\pi}\Phi^1_1 g^*; -\frac{g^*}{G_0  
  k_0^2}\!\right)\right]\\
  +\frac{1}{\alpha^*}\left( 1\!+\! \frac{g^*}{G_0 k^2}\right) {}_2 
  F_1\left(1,1,1\!+\!\frac{3}{\pi}\Phi^1_1 g^*; -\frac{g^*}{G_0 k^2}\right),
 \end{split}
\end{equation}
where ${}_2 F_1(a,b,c;z)$ denotes the (ordinary) hypergeometric function.

From eq.\,\eqref{aTrajectory2} we infer that there exist three kinds of 
possible UV behavior for $\alpha(k)$. They differ by the value of the terms 
inside the square brackets $[\cdots]$ on the RHS of \eqref{aTrajectory2}. This 
value is independent of the scale $k$. As the prefactor of $[\cdots]$ diverges 
proportional to $k^{6 \Phi^1_1 g^*/\pi}$ when $k\rightarrow \infty$, we find 
the limit $\lim_{k\rightarrow \infty}\alpha(k)= 0$ for every strictly positive 
value $[\cdots]>0$. This corresponds to asymptotic freedom of the 
fine-structure constant, and is similar to the behavior found by Robinson and 
Wilczek \cite{ymg-robwil} and Toms \cite{tomsnature}, but here with a 
concomitant running of Newton's constant. The corresponding RG trajectories of 
the full system are asymptotically safe with respect to {\bf NGFP}${}_1$.

For a negative value $[\cdots]<0$ there will be a scale $k_{\text{LP}}$ at 
which the two terms on the RHS of \eqref{aTrajectory2} cancel, so that $\alpha$ 
diverges at finite energies, corresponding to a Landau type singularity.

A third type of limiting behavior is obtained for the case that the bracket 
vanishes exactly: $[\cdots]=0$. As we are then left only with the second term 
of the RHS of \eqref{aTrajectory2}, and since ${}_2 F_1(a,b,c;0)=1$, we find 
$\lim_{k\rightarrow \infty}\alpha(k)= \alpha^*$, corresponding to an 
asymptotically safe trajectory with a non-zero coupling at the FP. This is 
precisely the behavior to be expected due to the UV repulsive direction of the 
fixed point. Since for $k\rightarrow \infty$ the trajectory will only flow into 
{\bf NGFP}${}_2$ for one specific value of $\alpha_0$, this value of 
$\alpha_0$, and hence the whole trajectory $\alpha(k)$, can be predicted under 
the assumption of Asymptotic Safety.
\begin{figure}
\centering
 \input{ASFlow-psfrag.tex}
 \includegraphics[height=8cm]{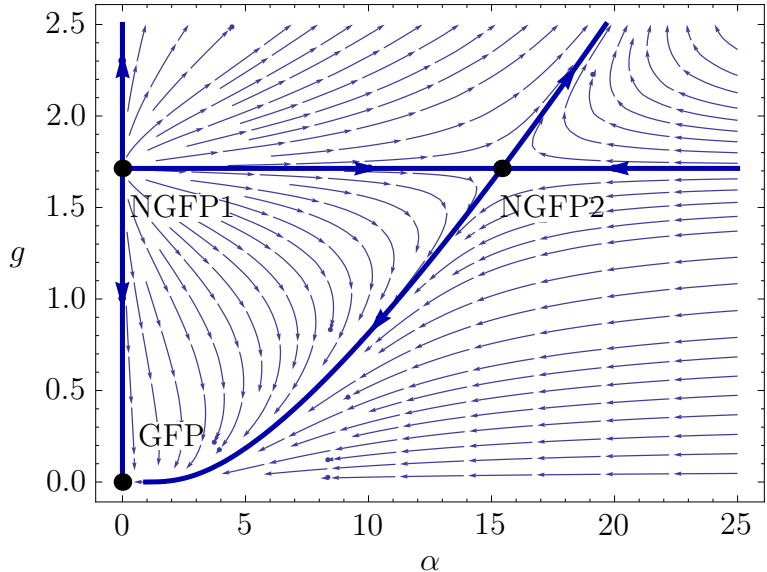}
\caption{The RG flow on the $g$-$\alpha$--plane implied by the simplified 
equations \eqref{FEapprox}. It is dominated by two non-Gaussian fixed points. 
Their respective value of $\dim \mathscr{S}_{\text{UV}}$ differs by one unit. 
(The arrows point in the direction of decreasing $k$.)}
\label{ASFlow}
\end{figure}

The situation is illustrated by the $g$-$\alpha$--phase portrait in 
Fig.\,\ref{ASFlow}. Bearing in mind that the arrows always point towards the 
IR, we see that ${\bf NGFP}_1$ is IR repulsive in both directions shown, while 
${\bf NGFP}_2$ is IR attractive in one direction. This is consistent with our 
earlier discussion which showed that in the 3-dimensional 
$g$-$\lambda$-$\alpha$--space ${\bf NGFP}_1$ has 3 and ${\bf NGFP}_2$ has only 
2 IR repulsive (or equivalently, UV attractive) eigendirections.

In Fig.\,\ref{ASFlow}, the trajectories inside the triangle ${\bf GFP}$--${\bf 
NGFP}_1$--${\bf NGFP}_2$ are those corresponding to the case $[\cdots]>0$ 
above; they are asymptotically safe with respect to ${\bf NGFP}_1$. The ${\bf 
NGFP}_2 \rightarrow {\bf GFP}$ boundary of this triangle is the unique 
trajectory (heading towards smaller $g$ and $\alpha$ values) which is 
asymptotically safe with respect to the second non-trivial fixed point, ${\bf 
NGFP}_2$.

The diagram in Fig.\,\ref{ASFlow} corresponds to a massless electron for which 
$A$ keeps its non-zero value at arbitrarily small scales. In reality the 
$\alpha$-running due to the fermions stops near $m_e$, of course.

\section{Asymptotic Safety construction at ${\bf NGFP}_2$}
Let us investigate the unique asymptotically safe trajectory emanating from 
${\bf NGFP}_2$ in more detail. First we note that the condition of a vanishing 
bracket $[\cdots]$ in \eqref{aTrajectory2} is self-consistent in the sense, 
that the resulting function $\alpha_0(k_0)$ is of identical form as the 
remaining function $\alpha(k)$:
\begin{equation}
 \frac{1}{\alpha(k)}= \frac{1}{\alpha^*}\left( 1+ \frac{g^*}{G_0 k^2}\right)\, 
 {}_2 F_1\left(1,1,1\!+\!\frac{3}{\pi}\Phi^1_1 g^*; -\frac{g^*}{G_0 k^2}\right).
\end{equation}
Second, let us approximate this function for scales $k \ll m_\text{Pl}$ much 
below the Planck scale. Later on we shall need it at $k=m_e$, for instance, 
where $m_e$ is the mass of the electron. Then the argument $\frac{g^*}{G_0 
m_e^2}=g^*\left(\frac{m_{\text{Pl}}}{m_e}\right)^2\approx 10^{44}$ is 
extremely large and this will be an excellent approximation. Hence we may 
safely truncate the general series expansion of the hypergeometric function,
\begin{equation}
 \begin{split}
  {}_2 F_1(a,a,c;z)=\frac{\Gamma(c)}{\Gamma(a) \Gamma(c-a)} &(-z)^{-a} 
  \sum_{n=0}^\infty \frac{(a)_n (1-c+a)_n}{(n!)^2} z^{-n} \cdot \\
  \cdot &\Big(\ln(-z) + 2\, \psi(n+1) - \psi(a+n) -\psi (c-a-n) \Big),
 \end{split}
\end{equation}
after its first term, and approximate the resulting factor $1+ G_0k^2/g^* 
\approx 1$, such that our final result for scales $k \ll m_{\text{Pl}}$ reads
\begin{equation}\label{ApproxResult}
 \frac{1}{\alpha(k)}= \frac{g^*}{\alpha^*} \cdot \frac{3}{\pi} \Phi^1_1 \cdot 
 \left[ \ln \left(\frac{g^*}{G_0 k^2} \right) - \gamma - 
 \psi\left(\frac{3}{\pi} \Phi^1_1 g^* \right)\right].
\end{equation}
Here $\psi$ denotes the Digamma function and $\gamma$ is Euler's constant. 
Using \eqref{FPValues} in order to reexpress the ratio $g^*/\alpha^*$ we can 
write \eqref{ApproxResult} also in the following form:
\begin{equation}
 \label{ApproxResult2}
 \frac{1}{\alpha(k)}= \frac{A}{2} \left[ \ln \left(\frac{g^*}{G_0 k^2} \right) 
 - \gamma - \psi\left(\frac{3}{\pi} \Phi^1_1 g^* \right)\right]
\end{equation}
Recall that $A\equiv \frac{2}{3 \pi} n_F$ is a completely universal constant, 
sensitive only to the number of (hypothetical) electron species. Hence, for 
$k\ll m_\text{Pl}$, we recover the logarithmic running $\alpha(k)^{-1}= - A 
\ln k + \text{const}$ familiar from pure QED.

In the opposite extreme of $k$ comparable to, or larger than the Planck mass 
the gravity corrections set in, stop this logarithmic behavior, and cause the 
coupling to freeze at a finite value $\alpha(k\rightarrow \infty)=\alpha^*$. 
Obviously, along this RG trajectory no Landau pole singularity is encountered!

Note also that according to eq.\,\eqref{ApproxResult2} we have $\alpha(k) 
\propto 1/n_F$ for every value of $k$. As a consequence, if we consider a toy 
model with a large number of electron flavors, all $\alpha$-values that appear 
along the RG trajectory can be made as small as we like, and this renders 
perturbation theory in $\alpha$ increasingly precise. At the fixed point we 
have for instance
\begin{equation}
 \alpha^* = 9 \, \Phi^1_1(0) \frac{g^*}{n_F}.
\end{equation}

In an Asymptotic Safety scenario based upon the fixed point ${\bf NGFP}_2$, 
within the truncation considered, the infrared value of the fine-structure 
constant $\alpha_\text{IR} \equiv \lim_{k\rightarrow 0} \alpha(k)$ is a 
computable number. Using eq.\,\eqref{ApproxResult2} to calculate 
$\alpha_\text{IR}$ we must remember however that as it stands it holds true 
only for $k\gtrsim m_e$. When $k$ drops below the electron mass the standard 
QED contribution to the running of $\alpha(k)$ goes to zero, and the gravity 
corrections are zero there anyhow. Hence approximately, $\partial_t 
\alpha(k)=0$ for $0\leq k\lesssim m_e$. Thus eq.\,\eqref{ApproxResult2} leads 
to the following prediction for $\alpha_\text{IR}\approx \alpha(m_e)$:
\begin{equation}\label{ApproxResult3}
 \frac{1}{\alpha_\text{IR}}= \frac{A}{2} \left[ 2\,  \ln 
 \left(\frac{m_\text{Pl}}{m_e} \right) + \ln(g^*) - \gamma - 
 \psi\left(\frac{3}{\pi} \Phi^1_1 g^* \right)\right].
\end{equation}
As the fixed point coordinates are an output of the RG equations, the only 
input parameter needed to predict $\alpha_\text{IR}$ in this approximation is 
the electron mass in Planck units, $m_e/m_\text{Pl}$.

It is tempting to insert numbers into eq.\,\eqref{ApproxResult3}. With $m_e= 
5.11 \cdot 10^{-4}\,\text{GeV}$ and $m_\text{Pl}=1.22 \cdot 10^{19}\, 
\text{GeV}$ one finds $m_e/m_\text{Pl}=4.19 \cdot 10^{-23}$, and for the 
optimized cutoff \cite{opt} we have $\Phi^1_1=1$. The value of 
$g^*=-2/B_1(\lambda_0)$ depends on the value chosen for $\lambda_0$. For 
$\lambda_0=0$ or $\lambda_0 = \lambda^*\approx 0.193$, the fixed point value of 
$\lambda$ in the Einstein-Hilbert truncation, we get $g^*\approx 1.71$ or 
$g^*\approx 0.83$, respectively. From that we obtain
\begin{equation}\label{Result1}
 \frac{1}{\alpha_\text{IR}}\stackrel{\lambda_0=0}{\approx} 10.91\, n_F \qquad 
 \text{or} \qquad 
 \frac{1}{\alpha_\text{IR}}\stackrel{\lambda_0=\lambda^*}{\approx} 10.96\,n_F.
\end{equation}
We observe that the result is relatively insensitive to the value of $g^*$ 
and/or $\Phi^1_1$, but it scales linearly with the number of electron species, 
$n_F$.

Obviously, for $n_F=1$, this estimate differs from the fine-structure constant 
measured in real Nature, $\alpha\approx 1/137$, by a factor of roughly 13. 
However, even within the limits of our crude approximation 
\eqref{Approximation}, a serious comparison with experiment must include the 
renormalization effects due to the other particles besides the electron, all 
those of the standard model, and possibly beyond. Within the ``$n_F$ flavor 
QED'' considered here we could mimic their effect by appropriately choosing 
$n_F$. It would then follow that the observed $\alpha_\text{IR}$ is 
consistent with Asymptotic Safety at ${\bf NGFP}_2$ if $n_F=13$. 

It is reassuring that for this large number the value of the natural expansion 
parameter of QED perturbation theory, $(\alpha/\pi)$, is rather small already. 
At ${\bf NGFP}_2$, for example, one has $(\alpha^*/\pi)\approx 0.38$ and 
$(\alpha^*/\pi)\approx 0.18$, respectively.

Next let us try the full Standard Model (SM) and its minimal supersymmetric 
extension (MSSM).\footnote{For a related discussion see \cite{WettShap}.} 
Applying the above discussion to the weak hypercharge rather than the 
electromagnetic $U(1)$ one again has a one loop flow equation of the type 
$\partial_t \alpha_1 = A \alpha_1^2$, this time with $A=41/(20 \pi)$ for the SM 
and $A=33/(10 \pi)$ for the MSSM, respectively \cite{amaldi}. Here $\alpha_1 
\equiv 5\alpha/(3 \cos^2 \theta_W)$ where $\theta_W$ is the Weinberg angle. It 
is most convenient to compare the prediction of Asymptotic Safety to the 
experimental value at the $Z$ mass. From eq.\,\eqref{ApproxResult3} with the 
new value of $A$ and $m_e$ replaced by $M_Z$ we obtain (with $\lambda_0=0$):
\begin{subequations}\label{prediction}
\begin{align}
 \alpha_1^{\text{SM}}(M_Z)&\approx 1/25.7\\
 \alpha_1^{\text{MSSM}}(M_Z)&\approx 1/41.3
\end{align}
\end{subequations}
As compared to the experimental value $\alpha_1^{\text{exp}} (M_Z) \approx 
1/59.5$ both of these predictions are too high, the supersymmetric one less so. 
Clearly we may not take these numbers too seriously. After all, while for the 
reasons discussed above we believe that the one loop form of the matter beta 
functions is a reliable guide with respect to the general structure of the RG 
flow, its quantitative status is questionable.

Nevertheless the following observation might be of interest. The predictions 
\eqref{prediction} turn out larger than the experimental value since in the SM 
and MSSM the coefficient $A$ is {\it too small}. As a consequence, the matter 
driven renormalizations which reduce $\alpha_1(k)$ when $k$ is lowered are too 
weak. If we could take the RG equations seriously at the quantitative level the 
conclusion would be that Asymptotic Safety at ${\bf NGFP}_2$ is possible if 
there exist more particles with a $U(1)$ charge than those of the SM or MSSM. 
We find it remarkable that not very many more seem to be needed; it is 
sufficient to increase $A$ by a small factor of order unity.

On the other hand, if ultimately it turns out that the standard model coupled 
to QEG is not asymptotically safe with respect to ${\bf NGFP}_2$ then its RG 
trajectory would be one of those inside the ${\bf GFP}$--${\bf NGFP}_1$--${\bf 
NGFP}_2$ triangle in Fig.\,\ref{ASFlow}. In this case it is asymptotically safe 
with respect to the other non-trivial fixed point, ${\bf NGFP}_1$. As for being 
free from divergences and predictive at all energies this is still not too much 
of a drawback, though. It only means that the $U(1)$ coupling is not a 
prediction but necessarily an experimental input.

\section{Numerical results}
Returning to QED coupled to QEG we shall now go beyond the analytically 
tractable approximation of the previous section and employ {\it exact} 
numerical solutions $g(k),\ \lambda(k)$ for the pure gravity subsystem of 
eqs.\,\eqref{FEg} and \eqref{FEl}. Thereby the exact form of the functions 
$\eta_N=B_1(\lambda) h_1(g,\lambda)$ and $\beta_\lambda(g,\lambda)$ as implied 
by the Einstein-Hilbert truncation \cite{mr} are used, and then the two coupled 
equations for $g$ and $\lambda$ are solved numerically as in \cite{frank1}. 
Then, for every given RG trajectory $k \mapsto (g(k), \lambda(k))$, we 
calculate the corresponding $\alpha(k)$ by inserting $g(k)$ into \eqref{FEa} 
and solving this decoupled differential equation numerically, too.

\begin{figure}
\centering
\subfigure[]{
 \includegraphics[height=6cm]{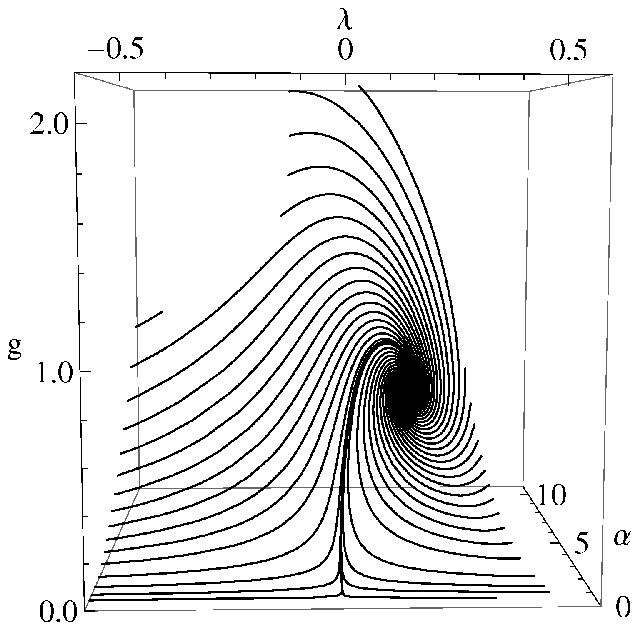}\label{SUV1}}\qquad
\subfigure[]{
 \includegraphics[height=6cm]{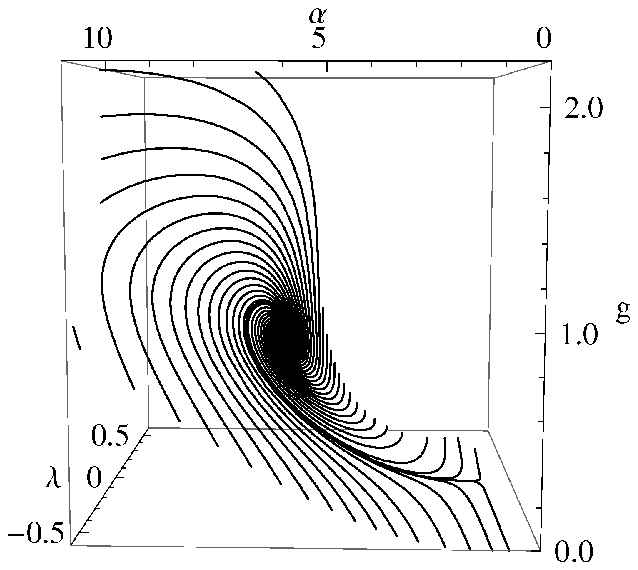}\label{SUV3}}\\
\subfigure[]{
 \includegraphics[height=11cm]{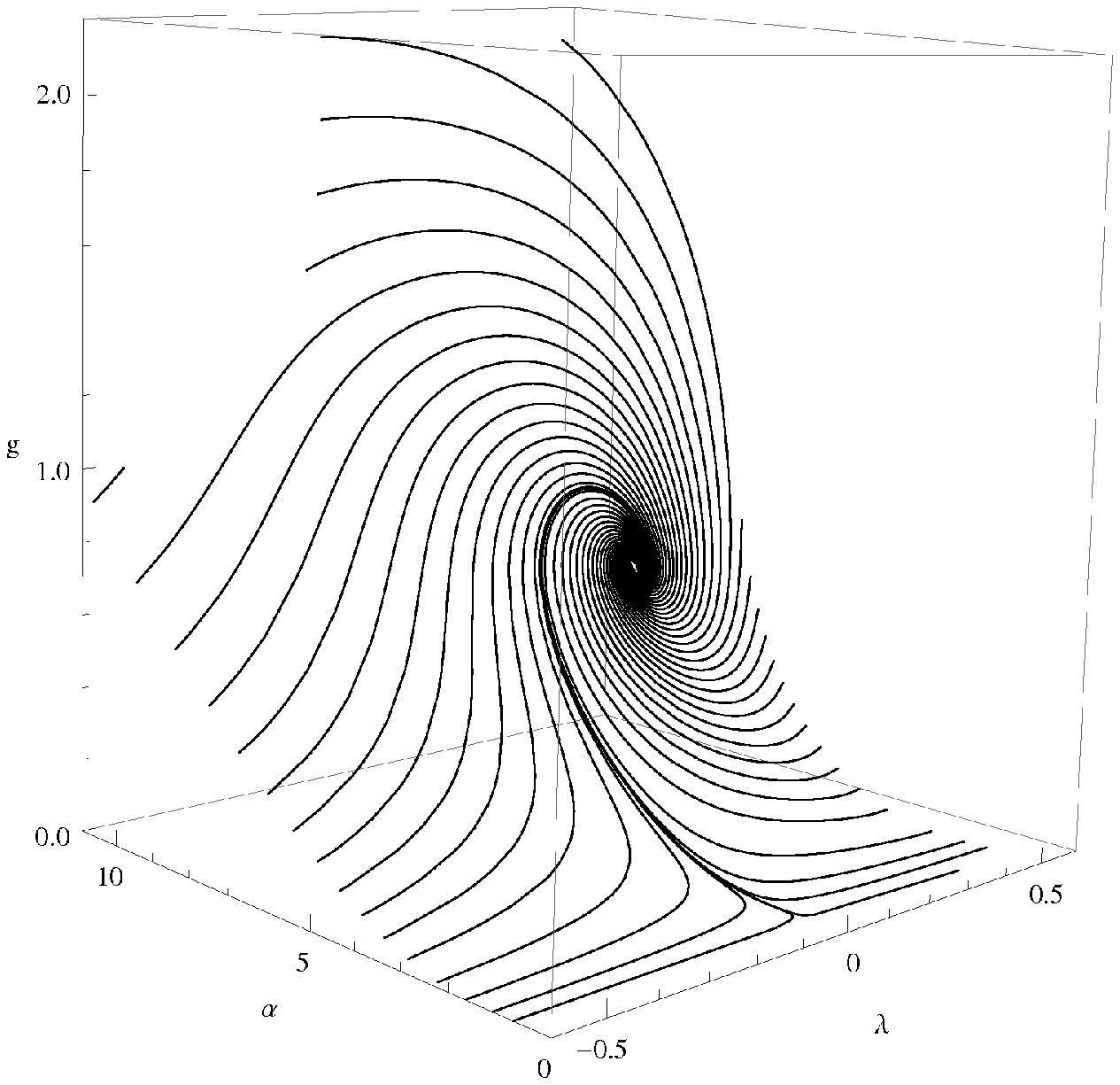}\label{SUV2}}\\
\caption{Trajectories running inside the UV critical surface 
$\mathscr{S}_{\text{UV}}$ of {\bf NGFP}${}_2$ in 
$g$-$\lambda$-$\alpha$--theory space.}
\label{SUV}
\end{figure}

Staying within the one-loop approximation of $\beta_\alpha$ we thus confirm the 
existence of both non-Gaussian fixed points, {\bf NGFP}${}_1$ and {\bf 
NGFP}${}_2$. In accord with the general discussion above, the latter is seen to 
have two UV attractive and one repulsive direction. All RG trajectories heading 
for $k \rightarrow \infty$ towards {\bf NGFP}${}_2$ lie in its two dimensional 
UV critical surface $\mathscr{S}_{\text{UV}}$. It is visualized in 
Fig.\,\ref{SUV} by a family of trajectories starting on 
$\mathscr{S}_{\text{UV}}$ close to the FP, which were traced down to lower 
scales $k$.

As the backreaction of the matter on the gravity sector is neglected, the flow 
in a projection onto the $g$-$\lambda$--plane is identical to the one of pure 
gravity (Fig.\,\ref{SUV1}). We can therefore classify the trajectories as in 
\cite{frank1}, being of type Ia, IIa, and IIIa, when the IR value of the 
cosmological constant is negative, zero, or positive, respectively.

As we rotate the coordinate frame (Fig.\,\ref{SUV2},\ref{SUV3}), we see how the 
critical surface is bent in coupling space. Especially we note that the 
fine-structure constant only gets renormalized to small values $\alpha \ll 1$, 
if the $g$-$\lambda$--projection of the trajectory is sufficiently close to the 
type IIa trajectory of pure gravity (the ``separatrix'' \cite{frank1}). This is 
because only these trajectories give rise to a long classical regime with 
$G,\Lambda\approx \text{const}$ \cite{entropy,h3}. They spend a tremendous 
amount of renormalization group time close to the Gaussian fixed point of the 
gravity sector. The classical regime of gravity is needed for the logarithmic 
running of $\alpha$ to be of effect.

As a concrete example of a trajectory with a long classical regime we consider 
the ``realistic'' RG trajectory discussed in \cite{h3} and \cite{entropy}. In 
these references a specific $g$-$\lambda$--trajectory has been identified which 
matches the observed values of $G$ and $\Lambda$. It is of type IIIa and can be 
characterized by its turning point (the point of smallest $\lambda$) whose 
coordinates are
\begin{equation}
 (g_T,\lambda_T)= \left(g_T, \frac{\Phi^1_2(0)}{2 \pi} g_T\right) \qquad 
 \text{with} \qquad g_T \approx 10^{-60}.
\end{equation}
The turning point is passed at the scale $k_T\approx10^{-30} m_\text{Pl}$. To 
make the numerical solution of the RG equations feasible we transform the 
equations to double logarithmic variables using $\tau(k) \equiv \ln(k/k_T) = 
\ln(k/m_{\text{Pl}})+30 \ln(10)$ as the RG time variable. The transition at 
the Planck scale between the classical and fixed point scaling regime therefore 
takes place at about $\tau(k=m_{\text{Pl}})=30 \ln(10) \approx 69$.

Having fixed the $g$-$\lambda$--trajectory to be the ``realistic'' one, there 
is a unique asymptotically safe trajectory relative to ${\bf NGFP}_2$ in the 
three dimensional coupling space. The corresponding $\alpha(k_T)$ can now be 
found by a shooting method: If we start slightly above 
$\mathscr{S}_{\text{UV}}\equiv \mathscr{S}_{\text{UV}}({\bf NGFP}_2)$ and 
evolve towards the UV the coupling $\alpha(k)$ will head to infinity at a 
finite scale, while starting below $\mathscr{S}_{\text{UV}}$ will result an 
asymptotically free trajectory: $\alpha(k\rightarrow\infty)=0$. This trajectory 
is asymptotically safe with respect to ${\bf NGFP}_1$, and in 
Fig.\,\ref{ASFlow} it corresponds to one of those {\it inside} the triangle 
${\bf GFP}$--${\bf NGFP}_1$--${\bf NGFP}_2$. The closer we get to 
$\mathscr{S}_{\text{UV}}$ the more the trajectory gets ``squeezed'' into the 
corner of the triangle at ${\bf NGFP}_2$, ultimately leading to two separate 
pieces, the ${\bf NGFP}_1\rightarrow{\bf NGFP}_2$ and the ${\bf 
NGFP}_2\rightarrow{\bf GFP}$ branch, respectively. Due to unavoidable numerical 
errors any starting value $\alpha(k_T)$ will eventually opt for one of the two 
cases, but if we fine tune it to happen at sufficiently large $\tau$, we will 
end up with a good estimate for the trajectory which is asymptotically safe 
with repsect to ${\bf NGFP}_2$, the boundary line ${\bf NGFP}_2 \rightarrow 
{\bf GFP}$ of the triangle in Fig.\,\ref{ASFlow}.

\begin{figure}
\centering
 \input{ShootingMethod-psfrag.tex}
 \includegraphics[height=6cm]{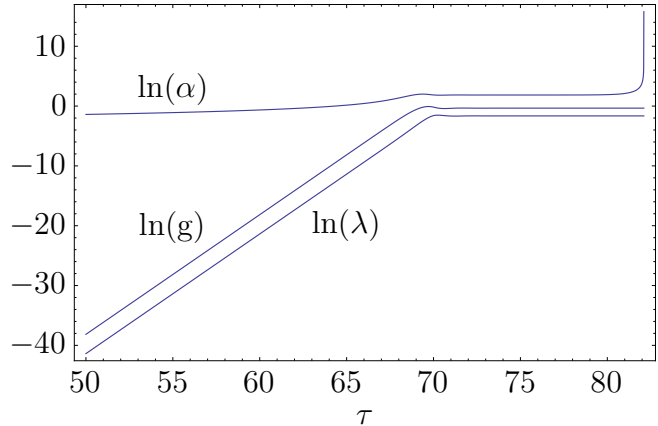}
 \caption{Double logarithmic plot of the running couplings of the ``realistic'' 
 trajectory.}
\label{ShootingMethod}
\end{figure}

The result of this procedure is depicted in Fig.\,\ref{ShootingMethod} where we 
set $n_F=1$.
As can be seen, there is a rapid transition to the fixed point scaling regime 
at the Planck scale ($\tau\approx 69$), above which all three dimensionless 
couplings remain constant at their fixed point values. By fine tuning we were 
able to choose the initial parameter to $1/\alpha(k_T)=14.65$ which ensures 
that the trajectory stays at the fixed point value for about two orders of 
magnitude in $k$ before it shoots away to infinity.

The scale of the electron mass corresponds to a $\tau$-value 
$\tau(k=m_e)=\ln(4.19\cdot 10^{-23})+30 \ln(10) \approx 17.55$, which is far 
above the turning point scale. At this scale the asymptotically safe trajectory 
predicts a value $\alpha(m_e)^{-1}\approx 10.93$ which is in perfect agreement 
with \eqref{Result1}. Hence we can conclude that the running of $\lambda$, as 
well as the exact functional form of $h_1(g,\lambda)$, are of little effect to 
the IR value of $\alpha$. 

\section{Summary and conclusion}

In this paper we coupled Quantum Electrodynamics to quantized gravity and 
explored the possibility of an asymptotically safe UV limit of the combined 
system. Using a simple truncation of the corresponding effective average action 
we found evidence indicating that this is indeed possible. There exist two 
non-trivial fixed points which lend themselves for an Asymptotic Safety 
construction. Using the first one, ${\bf NGFP}_1$, the fixed point value of the 
fine-structure constant is zero, and its infrared value $\alpha_{\text{IR}}$ 
is a free parameter which is not fixed by the theory itself but has to be taken 
from experiment. Basing the theory on the second non-Gaussian fixed point ${\bf 
NGFP}_2$ instead, the fixed point value $\alpha^*$ is non-zero, and the 
(``renormalized'') low energy value of the fine-structure constant 
$\alpha_{\text{IR}}$ can be predicted in terms of the electron mass in Planck 
units. In either case the coupled theory QED + QEG is well behaved in the 
ultraviolet, there is no Landau singularity in particular, and it is not 
trivial, i.\,e., the continuum limit is an (electromagnetically and 
gravitationally) interacting theory.

The key ingredient in the RG equations considered is the quantum gravity 
contribution to $\beta_\alpha$ which was obtained in ref. \cite{dhr1}. It is 
proportional to $g\,\alpha\equiv G(k) k^2\alpha$, and its sign is such that for 
increasing $k$ it counteracts the growth of $\alpha(k)$ caused by the fermions. 
This can lead to two qualitatively different scenarios for the high energy 
behavior of QED + QEG. In the first one, which is also seen in perturbation 
theory \cite{ymg-robwil, tomsnature}, the gravitational effects win over the 
fermionic ones and $\alpha(k)$ is driven to zero in the UV: it has become an 
asymptotically free coupling. For RG trajectories of this type the 
$k$-dependence of Newton's constant plays no essential role; the decrease 
of $G(k)$ becomes substantial only after $\alpha(k)$ is almost zero already. 
Instead, in the second scenario, the initial (low energy) value of $\alpha(k)$ 
is such that it has not yet become very small when the weakening of gravity due 
to the decrease of $G(k)$ sets in. In particular in the asymptotic scaling 
regime it decreases rapidly, $G(k)=g^*/k^2$, so that the fermions, still trying 
to increase $\alpha$ for $k\rightarrow \infty$, have a better chance to win 
over the gravitons now. Along certain trajectories they indeed do, but what is 
more interesting is the possibility of an exact compensation of the two trends. 
This is exactly what happens at the second non-trivial fixed point, ${\bf 
NGFP}_2$, which is characterized by a non-zero $\alpha^*$.

Note that this second possibility is closely related to the Asymptotic Safety 
of (pure) gravity. It could not be found in perturbation theory which, while 
using a similar gravity correction to $\beta_\alpha$, treats the factor of $G$ 
it contains as a constant and therefore misses the weakening of gravity at high 
scales.

The most remarkable feature of the second fixed point is the reduced 
dimensionality of its UV critical manifold and the resulting higher degree of 
predictivity than in perturbation theory. We take this as a first hint 
indicating that after coupling the standard model to asymptotically safe 
gravity it might perhaps be possible to compute some of its as to yet free 
parameters from first principles.\\

Acknowledgments: We are grateful to A. Ashtekar and H. Spiesberger for helpful 
discussions.

\end{document}